\documentclass{article}
\def\sla{\raise.15ex\hbox{$/$}\kern-.57em}
\begin{document}

\title{Exact Chiral Symmetry on the Lattice.
}

\author{Herbert Neuberger\\
Department of Physics and Astronomy, Rutgers University\\
Piscataway, NJ 08855\\
{\tt neuberg@physics.rutgers.edu}}

\maketitle

\begin{abstract}
Developments during the last eight years have refuted the folklore
that chiral symmetries cannot be preserved on the lattice. The
mechanism that permits chiral symmetry to coexist with the lattice
is quite general and may work in Nature as well. The reconciliation
between chiral symmetry and the lattice is likely to revolutionize the
field of numerical QCD. 
\end{abstract}

\vspace{-9.0cm}
\begin{flushright}{\normalsize RUHN-01-1}\\
\end{flushright}
\vspace*{8.0cm}

\section{INTRODUCTION}

Chiral symmetries play a central role in Particle and Nuclear Physics,
intimately related to such ubiquitous processes as $\beta$-decay. For
a while it seemed that one cannot find a non-perturbative regularization
of gauge theories with explicit chiral symmetry. There was a folklore
that chiral symmetry could not be realized on the lattice. This folklore
has been abandoned now. The  
breakthrough can be traced back to an early paper
by Callan and Harvey \cite{callan}, and to two subsequent papers, 
one by Kaplan \cite{kaplan} and
the other by Frolov and Slavnov \cite{slavnov}. 
Based on these papers a general way to
realize chiral symmetry on the lattice was devised and named the ``overlap'' 
\cite{ovplb1}.

Once the overlap construction was completed and it was established
that the required salient features of chiral fermions were indeed
successfully reproduced it was realized that one could have
arrived at the same construction not by starting from Callan and Harvey's
paper but rather from an older work of 
Ginsparg and Wilson \cite{ginsparg}. While the
Callan and Harvey paper pointed the direction towards chiral gauge
theories and was a natural outgrowth of the mathematical understanding
of anomalies that was developed in the early to mid eighties, Ginsparg
and Wilson focused on vector-like gauge theories and took a more 
traditional approach. It is a surprise that these two lines of thought
have merged. This review focuses on the overlap viewpoint because it
is the original way the construction was arrived at and because it is
simultaneously more physical and more modern mathematically. 

Another viewpoint can be found in the recent review 
of M. Creutz \cite{creutzrev}.
His review also contains a much more extensive list of references
and readers looking for more information than contained in the bibliography
of the present review are encouraged to consult this source. 

\subsection{Chiral Symmetry in Particle and Nuclear Physics}

In the minimal standard model the basic building blocks of matter are chiral
fermions. If we ignore weak and electromagnetic interactions, we are left
with QCD. QCD has almost exact global chiral symmetries and they explain
the low energy hadronic spectrum. The weak interactions come from gauged chiral
symmetries made consistent by delicate anomaly
cancelations ensuring their
preservation at the quantum level. Chirality continues to play a central role
in supersymmetric extensions of the minimal standard model. 
Grand unified theories, supersymmetric or not, still must be chiral gauge 
theories in order to reproduce the standard model. 

\subsection{The Effective Lagrangian Viewpoint} 

Relativistic field theory describes 
Particle Physics simply because it is a unique
structure incorporating special relativity, quantum mechanics
and the selfconsistency of a system consisting of a finite 
number of fundamental particles in interaction. 
Because of gravity, it is all but certain
that any field theoretical model is not more than an approximation
to Nature. This approximation tends to be very good in a certain energy
range. The top of this energy range can be viewed as the ``ultraviolet
cutoff'', $\Lambda$, of the theory. The larger $\Lambda$ is relative
to typical masses, $m_{\rm ph}$, in the theory, 
the more powerful and accurate 
is the effective model. Since the standard model describes 
Nature very accurately $\Lambda / m_{\rm ph}$
might be very large. 
However, for any theory to contain particles
of masses much lower than $\Lambda$, one needs a specific 
mechanism; without one, it is just too unlikely that 
$m_{\rm ph} << \Lambda$. The larger
the $\Lambda / m_{\rm ph}$ is, the more awkward
is the situation in the absence of a special mechanism. 
Luckily, mechanisms
that produce very light or even massless particles are known 
for spin 0, 1/2 and 1: Spin zero particles are naturally light 
if they are Goldstone
bosons associated with the spontaneous breakdown of a global symmetry.
Spin 1 particles are naturally massless if they are the force carriers
in a gauge interaction. Spin 1/2 particles are naturally massless if
they are chiral. This takes care of all particles except the yet 
undiscovered Higgs. One way to also make the Higgs naturally light,
although it is not a Goldstone boson, is provided by supersymmetry. 
Supersymmetry places particles of different spin into
equal mass (super)multiplets, so the above mechanisms can
also restrict the mass of particles of the ``wrong'' spin. 

One indication for what  $\Lambda$ might be is given by
the confluence of the three known gauge couplings at a high energy. 
There may exist a very accurate ``Grand Unified'' (GU)
effective field theory description of Nature with 
a $\Lambda \sim 10^{16} GeV$! 
Chiral symmetries continue to play a central role in GU theories,
be they supersymmetric or not.

\section{CONTINUUM}

Chiral fermions in interaction with gauge fields 
have some peculiar properties in the continuum, at the semiclassical
level \cite{contin}. These properties are best expressed in modern mathematical 
terms. They all deal with chiral fermions in the background provided by
a gauge field that is smooth up to gauge transformations. 
The fermion fields do not need to
be smooth; their fluctuations are easily brought
under control because their equations of motion are linear. 
Physically, one expects that fermionic and bosonic fluctuations
are controlled by the same ultraviolet cutoff
and the quantum fluctuations in the 
gauge background also have to be
taken into account.
Perturbatively it is known how to do this, but if the interactions
are strong the situation is less certain. Much of this uncertainty has been
eliminated by the developments central to this review, but without
a good grasp of the continuum semiclassical and 
perturbative properties neither
the problems nor their resolution can be fully appreciated.

The gauge invariant content of a gauge configuration defines a gauge
orbit. The multitude of gauge orbits forms a space that has an infinite
set of disconnected components, each uniquely identified by a signed integer,
$n_{\rm top}$ the ``topological charge''. 
This is best understood in the Euclidean 
formulation of gauge theories, where space-time is replaced by a four 
dimensional compact manifold. For any fixed gauge background one can
calculate expectation values of strings of fermion field operators.
Strangely, it turns out that single (or odd numbers of)
fermion operators can have non-vanishing expectation values in certain
cases, when $n_{\rm top}\ne 0$ \cite{thooft}. This is surprising 
because in a literally interpreted path
integral only products
of even numbers of Grassmann variables can integrate to non-zero values. 

When chiral fermions are integrated out one expects to
get a functional of the gauge fields given by the determinant of the
Weyl operator. Under a gauge transformation this operator transforms
by conjugation, but, sometimes, the determinant unavoidably breaks gauge
invariance. This is the single source of anomalies: the expectation value
of any fermionic observable transforms as expected under a gauge 
transformation, up to the chiral determinant which enters as a multiplicative
factor. Only the phase of this multiplicative
factor can have anomalies.
There is therefore no way one can view the chiral determinant as
a function on orbit space, because its phase may have to vary along
some orbits. There is a mathematical construct which usefully
accommodates this situation: One associates with each orbit
a one dimensional complex vector space and the collection of these
spaces makes up a smooth manifold. The chiral determinant is a section
in this ``line bundle'', associating one vector to each orbit. The
information contained in this vector is that of one complex number since
the vector space is one dimensional. 
The determinant line bundle may be twisted
in the sense that it cannot be parameterized globally and
smoothly as a direct product
of orbit space times a complex line. When there are twists, all
sections must have some zeros. In turn, these zeros are related
to anomalies. Thus, 
in the case of compact gauge groups, anomalies have a definite
geometric origin which explains why they are irremovable and why they
are associated with quantized numerical coefficients. 
In addition, orbit space is relatively complicated, and can
contain non-contractible loops around which the chiral determinant can
have nontrivial monodromies leading to so called global anomalies. 
 
All in all, it is better not to force 
the determinant to be a function over orbit space, but define it first
as a line bundle over orbit space. 
Avoiding to
think about the chiral determinant as a function is compatible with
the Weyl operator being a map between spaces that naturally carry
distinct spinorial representations of the Euclidean rotation group.
There is no natural identification between these spaces, so it
is better to stay with a view of the chiral determinant as a map
between some derived spaces. This problem disappears when one considers
the product of the Weyl operator with its adjoint and this is compatible
with the absolute value of the chiral determinant being a naturally
defined function over orbit space. 
 
There exist mathematical relationships between anomalies in different
dimensions, with distinct roles played by the even and odd dimensional 
subsets \cite{descent}. 
These relationships are described by so called ``descent equations''
which can be physically realized by considering defects of dimension
one or two embedded in a manifold of higher dimension \cite{callan}. 

Anomalies can also be understood in the context of local cohomology 
\cite{cohom}
as breaking gauge invariance by closed but not exact forms that are
locally expressed in terms of the gauge field. Both local and global cohomology
considerations also enter in the BRS formulation of gauge invariance.

These continuum properties indicate that path integrals should not be taken
too literally when one deals with chiral fermions. Nevertheless,  
the needed generalization is very mild and path integrals work
almost perfectly. 

\section{LATTICE DIFFICULTIES}

In lattice field theory one replaces the compact Euclidean manifold by
a finite grid of points. Almost all work is restricted to tori. The torus
is replaced by a regular lattice preserving a large discrete subgroup 
of translations. The inverse of the nearest neighbors spacing $a$ in the grid
is a physical ultraviolet cutoff and, at 
distances much larger than $a$ an effective
continuum Lagrangian description becomes valid. This effective Lagrangian,
so long it is used only in the regime where it applies, can work then
just as well as the true effective Lagrangian of Nature does,
although nobody thinks that spacetime really is a lattice. The
lattice construction is made outside perturbation theory and this is its
main advantage. For the lattice to be useful one needs some mechanism on the
lattice that would ensure exact chiral symmetry in the effective Lagrangian.
A finite grid implies a finite Grassmannian path integral which is well
defined and therefore no longer has the flexibility to admit the needed
slight generalization mentioned in the previous section. This reflects
itself in some difficulties when trying to enforce chiral symmetry on the
lattice. In addition, orbit space on the lattice is connected so
before one speaks about strange effects at nonzero $n_{\rm top}$ one
needs to give a meaning to $n_{\rm top}$. To successfully take
chirality to the lattice one must also take gauge field topology to the
lattice and do this in a selfconsistent way. 

\subsection{Nielsen-Ninomiya Theorems}
The lattice is unique as an ultraviolet cutoff in that it can
preserve gauge symmetries. Given a lattice action with finite 
interaction range and some global internal symmetry it is guaranteed
that one can gauge that symmetry, producing a new exactly gauge invariant
action. Let us assume that we can write down a finite range lattice
field theory describing a single Weyl fermion. Since we could gauge
its $U(1)$ symmetry and the resulting theory ought to be anomalous we
reached a paradox. The resolution to the paradox is that the toroidal
structure of lattice momentum space implies in this case that there
will be other massless excitations, which, after gauging, 
compensate the anomaly \cite{nielsen}.
In particular, QCD with a fermion action invariant
under chiral transformations (i.e. massless quarks) cannot be put on
the lattice because of the absence of obstacles to gauge its
chiral symmetries. 

\subsection{Instantons and Undesired $U(1)$'s}

Instantons imply that some of the global symmetries of
a fermionic action are destroyed by quantum effects. This is why
there is no ``ninth Golstone boson'' in QCD. However, a global
symmetry of a local lattice action is indestructible \cite{eichten}. 

\subsection{Troubles with BRS Invariance}

In the continuum one can replace the principle of gauge invariance
by that of $BRS$ symmetry \cite{brs}. This might indicate that one could 
restrict ones attention to gauge fixed actions avoiding the head
on collision with the lattice's ability to easily accommodate
gauge invariance. 
Unfortunately, it seems that $BRS$ symmetry does
no generalize to the lattice in an exact way \cite{brsnogo}. 

\section{HEURISTIC DESCRIPTION OF NEW APPROACH}

Lattices and chiral symmetry have been reconciled
by slightly enlarging the concept of fermion path integration in
that infinite numbers of fermions are allowed to live on the 
traditional finite grids. 

\subsection{Infinite Number of Fermions}
If we do not insist on chiral symmetry one can easily 
put (massive) Dirac fermions on the lattice. 
The known lattice difficulties imply that the chirality
breaking terms cannot be removed without changing the theory
in a substantial way, which goes far 
beyond rendering the Dirac fermions massless. 
For several Dirac fermions we can introduce a general
mass matrix, but unitary transformations of the fermion fields
can be used to make this matrix diagonal with non-negative entries.
However, if we make the mass matrix infinite and view flavor space
as an infinite dimensional Hilbert space, we can arrange that the
above ``unitary'' operations be not allowed. Moreover, choosing the
mass matrix to be a Fredholm operator with nontrivial index
we can destroy the balance between the left and right Weyl 
components of the Dirac fermions. One of the ``unitary'' operators
that would normally rotate one of the fermion fields to produce a diagonal
mass matrix becomes a partial isometry, and therefore no longer
induces an innocuous change of variables in the fermion integral. 
In short, the generalization of path integrals we require is that
it admit a certain kind of infinite flavor space. In Feynman diagrams
one only needs to make sure that all infinite sums one encounters
converge. Some transformations of integration variables in the path
integral are no longer allowed. The path integral itself no longer
is guaranteed to give an unambiguous answer, since the matrix whose
determinant we would need to take is infinite \cite{ovplb1}.

\subsection{Infinite Mass Matrix from Domain Walls}
The basic idea described above was motivated by
the domain wall setup \cite{kaplan}. With the benefit of hindsight
the logical order can be reversed and domain walls can be viewed 
as one particular realization of a mass matrix 
operator with index.
 
\section{FUNDAMENTALS}
The basic idea is to employ a Lagrangian whose fermionic
terms have the following structure:
\begin{equation}
{\cal L}_\psi = \bar\psi\sla D\psi +
\bar\psi(P_L {\cal M}+P_R {\cal M}^\dagger )\psi, ~~~
P_L={1\over 2} (1+\gamma_5 ), P_R={1\over 2}
(1-\gamma_5).
\end{equation}
The fermionic fields are Grassmann valued states in the flavor Hilbert
space, indexed by a spacetime coordinate, a Dirac index and a group
representation index. On a lattice the spacetime coordinate becomes
a discrete index of finite range. The mass matrix is an operator
in flavor Hilbert space and flavor index is summed over by viewing
each term in the Lagrangian as an inner product in flavor Hilbert space. 
This is possible because the action is bilinear. The main property
of the mass operator is that it has unit index \cite{ovplb1} 
and that it is
proportional to a dimensionful mass parameter which plays the role
of an ultraviolet cutoff in an effective theory
working at much lower energy. Chirality is not at all necessarily apparent 
in the full theory, but emerges in the effective description of low
energy physics in a natural way, as a consequence of the structure
of the mass matrix in flavor space. If Nature works in a similar way,
ordinary chirality has no fundamental meaning. The full
theory has a new mechanism which makes chirality appear naturally
at low energies. 

\subsection{Lattice Transfer Matrix and Overlap}

A simple choice for the flavor Hilbert space is that it be the
linear space of infinite complex sequences with $L_2$ inner product.
These sequences can be viewed as living on a one dimensional 
infinite lattice. It is easy to write down a simple bounded operator
with index acting in this space. In the site basis it reads:
\[
{\cal M}_{s,s^\prime} =\delta_{s+1,s^\prime} 
-a(s)\delta_{s,s^\prime},~~~s,s^\prime \in Z ,
\]
\begin{equation}
a(s)=a_+^0~~{\rm for}~s\ge 0, ~|a_+^0|<1,~~~
a(s)=a_-^0~~{\rm for}~s <0, ~|a_-^0| >1.
\end{equation}

One needs to be specific about the form of the other terms in the
fermionic Lagrangian, this time specializing to the lattice. 
\[
{\cal L}_\psi ={1\over 2} \bar\psi \gamma_\mu
(T_\mu - T_\mu^\dagger ) \psi +
\bar\psi P_L {\cal M}(w) \psi +\bar\psi P_R {\cal
M}^\dagger (w) \psi,
\]
\begin{equation}
T_\mu (\psi) (x) =
U_\mu (x) \psi(x+\mu ),~~~w=\sum_\mu (1- {{T_\mu+
T_\mu^\dagger }\over 2}).
\end{equation}

To avoid fermion doubling the mass operator must act 
nontrivially also on indices other than flavor.
${\cal M}$ acts trivially on spinor indices.
In the absence
of gauge fields, we need to make the parameters $a_\pm$ in
${\cal M}$ operators in spacetime in such a way that the index
of ${\cal M}$ be unity only in a small region around the
origin of momentum space, but vanish outside this region.
Thus, we avoid creating massless fermions at the doubler
locations. It is crucial that this can be accomplished
with a ${\cal M}$ that is totally smooth in its dependence on
momenta. Clearly, the spectral properties of ${\cal M}$ are
not smooth in their dependence on momenta, and this is possible
only because ${\cal M}$ is an operator in an 
infinite dimensional space. When the gauge field is turned on,
gauge covariance dictates that the parameters $a_\pm$ also
act now on the representation index:

\begin{equation}
a_\pm = a_\pm^0 +w.
\end{equation}

The next step is to make sense of the fermionic path integral in
this setting. This is done by introducing a finite dimensional
fermionic Fock space generated by creation and annihilation operators
$\hat a^\dagger, \hat a$ acting on a vacuum. There is one new
pair of $\hat a^\dagger, \hat a$ for each lattice site $x$, group index $i$
and Dirac index $\alpha$. 

Associated with the two semi-infinite lines
$s>0$ and $s<0$ there are two ``evolution'' operators, 
${\hat T}_\pm = e^{-\hat a^\dagger H_\pm {\hat a}}$, 
where $T_\pm=e^{-H_\pm}$ are transfer matrices. 
The $\hat T_\pm$ are smoothly parameterized
by the link variables $U_\mu(x)$. The 
path integrals over the fermion fields
located on the semi-infinite lines are interpreted 
as projectors on the ground states
of $\hat T_-$, $\hat T_+$ so the entire
path integral produces the overlap formula
for the lattice regulated chiral determinant \cite{ovnpb1}:
\begin{equation}
{\rm chiral ~determinant=}~\langle -\vert +\rangle .
\end{equation}

\subsection{Continuous Flavor Space}

Our construction did not make use of the boundedness of the
operator ${\cal M}^\dagger {\cal M}$, so we may as well make
flavor space continuous and enjoy the simplified form of
the single particle Hamiltonian matrices \cite{ovnpb2}:
\begin{equation}
H_\pm=\gamma_5[m_\pm +\sum_\mu (1-V_\mu )],~~~
V_\mu=(V_\mu^\dagger )^{-1}=
{{1-\gamma_\mu}\over 2}T_\mu +
{{1+\gamma_\mu}\over 2} T_\mu^\dagger .
\end{equation}
The mass parameters are limited by $m_- >0$
and $-2 < m_+ <0 $. Nothing of importance changes
in the limit $m_-\to \infty$ \cite{boya}. Overall scales
of $H_\pm$ do not enter so we can take 
$H_- =H^\prime=\gamma_5$ and $H_+ = H$. The main point
is that $H$ provides a lattice description of
Dirac fermions of negative mass, while $\gamma_5$
describes Dirac fermions of positive (actually infinite) 
mass. The matrix $m+\sum_\mu (1-V_\mu)$
is nothing but the Wilson-Dirac operator $D_W$;
it has no chiral symmetry, but by tuning $m$
to a gauge coupling-dependent value one can
recover a single massless Dirac fermion in 
the continuum limit. It is important to keep
in mind that in $H_\pm$ the mass parameters
are not to be tuned.

For the ground state of $\hat a^\dagger H \hat a$
to be non-degenerate we need to know that $H^2$
is bounded away from zero, something we shall assume 
and come back to later. One can rescale
$H$ not only by a constant but also by any 
positive function of $|H|$ which leads to
replacing $H$ by $\epsilon = {\rm sign} (H)$.
To unify notation, we denote $\gamma_5 =\epsilon^\prime$. 
$\epsilon^\prime$ and $\epsilon$ are reflections, and
associated with them are hermitian projectors 
$P={1\over 2}(1-\epsilon)$ and $P^\prime=P_R$. Let $N$
be the total dimensionality of the space in which these
matrices act. The projectors define two orthogonal
decompositions of the total space: the one
associated with $P^\prime$ is standard and gauge field
independent and the other depends on the gauge 
field smoothly, so long the sign function $\epsilon$ is defined.
The non-trivial subspaces are spanned by $\vec v_i$
with $P {\vec v}_i = \vec v_i,~i=1,2,...N_v$ and by
$\vec w_i$ with $P {\vec w}_i = 0,~i=1,2,...N-N_v$. The
trivial subspaces are associated with $P^\prime$, and the
related vectors are primed. It is convenient to introduce
$N\times N_v$ matrices $v=(\vec v_1 , \vec v_2, ....,\vec 
v_{N_v})$ and do the same with the other collections of
vectors. Then $P=vv^\dagger,~~v^\dagger v=1,~~1-P=ww^\dagger$. 
If $N_v=N/2$ the overlap is
\begin{equation}
\langle v^\prime \vert v \rangle = \det (v^{\prime\dagger}v),
\end{equation}
and if $n_{\rm top} \equiv N_v -N/2$ does not vanish, 
$\langle v^\prime \vert v \rangle =0$ \cite{ovnpb2}.

\subsection{Effective Low Energy Theory}

The effective theory describing only the massless
fermion modes is directly obtainable from the propagators
in the full theory, because the fermions enter bilinearly. 
Actually, we don't even need the action if we have the
propagators and the determinant, and, in addition, know
how to handle zero modes. Thus, we can eliminate from
the picture the entire flavor space, making the domain
walls disappear. 

Let us first assume that $n_{\rm top}=0$. Then
the propagator is obtained from the matrix
element of the creation/annihilation operators
between the two vacua \cite{ovnpb2}:
\begin{equation}
G^R_{JI} = {{\langle v^\prime \vert \hat 
a_I^\dagger \hat a_J \vert v\rangle}\over
{\langle v^\prime \vert v\rangle}} \Rightarrow 
G^R = 
v [v^{\prime\dagger} v]^{-1}v^{\prime\dagger}.
\end{equation}

The matrices $v$ and $v^\prime$ are defined up
to unitary transformations $v\rightarrow v U$
$v^\prime\rightarrow v^\prime U^\prime$. These
transformations change the phase of 
$\langle v^\prime \vert v\rangle$ but leave
$G^R$ invariant. Thus, the ambiguity exploited
by potential anomalies is restricted to affecting
the chiral determinant only. Also, it
is clear that 
\begin{equation}
w^\dagger G^R =0,~~~G^R w^\prime=0,
\end{equation}
showing that $G^R$ is of reduced rank, as it 
should be since it represents Weyl fermions
in a space large enough to accommodate Dirac
fermions. 

There is an additional freedom in the definition
of $G^R$ stemming from the ordering of $\hat a^\dagger$ 
and $\hat a$. A more symmetric 
definition is possible: $G^R \rightarrow G^R
-1/2$. With it, defining $G^L$ from the matrices
$w^\prime$ and $w$ and again shifting, the
new propagators satisfy $G^R=-G^{L\dagger}$,
just like in the continuum.

So long $n_{\rm top}=0$ only products containing
equal numbers of $\hat a^\dagger$ and $\hat a$
have non-zero matrix elements, and these are 
determined completely by the bilinear we just
evaluated as a consequence of a 
slight generalization of Wick's theorem. 
When $n_{\rm top}\ne 0$, to get a non-zero 
matrix element one needs a product containing 
unequal numbers of creation and 
annihilation operators, the imbalance being 
determined by $n_{\rm top}$. It is not hard
to derive explicit formulae for this case too \cite{ovnpb2}.

The collection
of second quantized state $\vert v\rangle$ as a 
function of the gauge background makes up a line
bundle over the space of link variables
for which the sign function is defined.  
The line bundle of the $\vert v^\prime\rangle$ is 
trivial and gauge invariant. This
makes the overlap a line bundle over orbit space,
just the kind of solution to the problem
of chiral fermions the results in the 
continuum were pointing to. The $\vert v\rangle$
line bundle descends from the $N_v$ dimensional
vector 
bundle of subspaces defined by $v$. The latter
bundle has a natural complement in the bundle
defined by $w$. The sum of these bundles is
obviously trivial, so the twists in each
are complementary, explaining the tight relation
between right and left handed Weyl fermions.
In particular, $\langle w^\prime \vert w\rangle^*
=\langle v^\prime \vert v\rangle$ \cite{ovnpb2}.

\subsection{Perturbative Calculations}

To check the formulation in perturbation theory one can
proceed in two ways. The quicker way is to calculate
the chiral determinant numerically for simple perturbative
gauge backgrounds, take the continuum limit and compare
to the perturbative continuum result \cite{ovnpb2,prl1}. The success of such
calculations is strong indication that perturbation theory
works. Analytical calculations take more effort and time, 
but they leave no doubts.

Several types of perturbative calculations have
been carried out. The first was an analytical calculation of the gauge 
anomaly in two dimensions for a chiral fermion in a $U(1)$ gauge
field \cite{ovnpb1}. Soon after, this calculation was followed by numerical
calculations in special backgrounds in two and four dimensions,
all concentrating on the anomaly and getting the 
correct consistent coefficient in the abelian case \cite{ovnpb2,prl1}.

The first comprehensive set of analytical calculations were carried
out by S. Randjbar-Daemi and Strathdee \cite{daemi}. All chiral anomalies
were calculated in the overlap and shown to come out right, and even Higgs
fields were included. It was established that the minimal standard model
in the overlap formulation worked correctly to one loop order.
This included calculations of the vacuum polarization and of RG functions.
Also, both the covariant and the consistent forms of the anomalies
were shown to come out correctly. An important extensions of the formalism
was the calculation of Lorentz anomalies for chiral
fermions coupled to two dimensional Euclidean gravity. 

A new wave of perturbative calculations \cite{newpert}
focused on the vector-like
context, where the overlap simplifies in a manner to be discussed later on.
Earlier calculations focused on the abelian axial anomaly. More
difficult calculations were carried out for the lattice $\Lambda$-parameter
and the renormalization of various operators, quark bilinears and
four quark operators. These calculations also dealt with currents.
This work will be essential in interpreting numerical QCD data obtained
using overlap fermions and extracting phenomenological numbers from
it. 

Surprisingly, it turned out that an old perturbative calculation of
Ginsparg and Wilson implied that the overlap has the right axial
anomaly. In the context of domain walls there were early
calculations of the covariant axial anomaly \cite{kaplan} 
and later calculations
of one loop corrections to the quark mass when the chiral symmetry
is only approximate \cite{kikukawa}. 
More recently several domain wall calculations
were carried out with a QCD orientation \cite{aoki}.

\subsection{Odd Dimensions}

In odd Euclidean dimensions masslessness of 
fermions is protected by parity \cite{mooregaume}. 
There are no perturbative anomalies and 
no topology
induced expectation values of unbalanced
products of $\bar\psi$ and $\psi$. 
There is less topological structure
in the space of gauge orbits. The single
kind of problem one may encounter has to do
with global gauge anomalies, which
occur only if one insists on parity invariance.
The overlap, as a mechanism
for making fermions massless, can be used 
just as in even dimensions, only now we should
expect that a natural section choice in the line
bundle exist, so long we relax parity requirements.

By embedding the odd dimensional
torus in a degenerate one of one dimension higher
we can use the even dimensional overlap to
define the odd dimensional one \cite{kikneu}. 
Making some global basis choices
one learns that the overlap hamiltonians in
odd dimensions have a simple structure
\begin{equation}
H = \pmatrix { 0 & D_W \cr D_W^\dagger & 0},
~~~H^\prime = \pmatrix {0 & -1\cr 1& 0},
\end{equation}
where $D_W$ is the Wilson-Dirac operator in
odd dimensions $d$, using one of the two inequivalent $2^{{d-1}\over 2}$ 
dimensional representation of the 
$\gamma$-matrices. We are assuming that
the gauge background is such that $D_W$ is
invertible. The following identity is derived
by manipulations familiar from finding solutions
to the Dirac equation in the continuum:
\begin{equation}
H\pmatrix { V {\vec \xi}_i \cr -{\vec \xi}_i}
= -\pmatrix
{\sqrt{D_WD_W^\dagger} & 0\cr
0 & \sqrt{D_W^\dagger D_W}} \pmatrix 
{V{\vec \xi}_i \cr -{\vec \xi}_i }.
\end{equation}
$V$ is a unitary matrix,
\begin{equation}
V= D_W{1\over\sqrt{D_W^\dagger D_W}} =
{1\over\sqrt{D_W D_W^\dagger}} D_W
\end{equation}
and ${\vec \xi}_i$ 
is an arbitrary normalized vector.
There are $N/2$ independent vectors 
${\vec \xi}_i$ and
the above identities show that they can 
build a basis of the negative
eigenspace of ${\rm sign} (H)$, trivializing
the vector bundle of negative eigenspaces over
gauge backgrounds. Let $\xi=({\vec \xi}_1,
{\vec \xi}_2,....{\vec \xi}_{N/2})$.
Similarly, we get a gauge field independent
basis for the negative eigenspace of 
${\rm sign} (H^\prime )$, $\xi^\prime$ 
producing
\begin{equation}
v^{\prime\dagger} v = \xi^{\prime\dagger}
{{1+V}\over 2} \xi .
\end{equation}
$\xi$ and $\xi^\prime$ are gauge field 
independent and can be chosen to 
be $N/2\times N/2$ unit matrices. Then,
$v^{\prime\dagger} v ={{1+V}\over 2}$. 
This expression is obviously gauge invariant. 
It is easy to also derive explicit expressions
for the fermionic propagators. 
The odd dimensional descendants of the
other handedness correspond to the 
other inequivalent representation
of odd dimensional $\gamma$-matrices. 

In odd dimensions one can defined parity as 
a flip of all coordinates. If the gauge field
in $V$ is replaced by its parity transform
$V$ changes to $V^\dagger$. We see therefore
that the fermion determinant conserves parity
only if $\det (V) =1$, and parity will end up
necessarily sacrificed when the potential for
global anomalies arises. Note that 
$\det (V)$ is the exponent of a local gauge 
invariant functional of the gauge field
\begin{equation}
\det (V)= \det (D_W) / \vert \det (D_W) \vert ,
\end{equation}
so that parity could be restored if the spectrum
of $V$ could be shown to have a gap (for all 
gauge fields) at some point on the unit circle.
Then, it would be possible to define smoothly
$\sqrt{\det (V)}$ and use this quantity to
redefine the overlap phase, so that both
gauge invariance and parity be preserved. 
If there is a global anomaly one can show that
a gap at a fixed location on the unit circle
is excluded because some loops of gauge orbits
make $\det (V)$ wind round the circle. 

We learn that the overlap is flexible enough to
work across all dimensions. Also, there is great
benefit in exploiting the relations between consecutive
dimensions, as indicated by experience in the
continuum.

\subsection{Spectral Properties of $H$}

For the sign function to be defined we need
to avoid situations where $H$ has an exact zero mode.
This can be assured by a local, 
gauge invariant constraint on the gauge fields:
The spectrum of $H$ clearly is gauge invariant,
so any condition on it is gauge invariant too.
For a trivial background the matrices $T_\mu$ 
can be simultaneously diagonalized, 
and the spectrum of $H$ has a finite gap 
around zero. By continuity, this gap cannot close
when the matrices $T_\mu$ are slightly 
deformed by turning on a nontrivial background, 
so that they no longer commute.
The norms of all $T_\mu$ commutators are 
gauge invariant and 
controlled by the norms of the plaquette 
variables, because the $T_\mu$'s are elementary
parallel transporters. 
When the plaquette variables are close to 
unity the commutators are small in norm and the gap is present.

This can be made rigorous and one can prove 
the following inequality about the 
smallest eigenvalue $\lambda_{\rm min}$ \cite{neubound}:
\begin{equation}
\left [ \lambda_{\rm min} \left (
D_W^\dagger D_W \right ) \right ]^{1\over 2}
\ge \left [1-(2+\sqrt{2}) \sum_{\mu > \nu } 
\epsilon_{\mu\nu}\right ] - |1+m|.
\end{equation}
$-2<m<0$ is the mass parameter, and
every plaquette variable $U_{\mu\nu(x)}$
obeys
\begin{equation}
\| 1-U_{\mu\nu}(x)\|\le\epsilon_{\mu\nu} ,
\label{plaqbnd}
\end{equation}
So long the $\epsilon_{\mu\nu}$'s are small 
enough a gap is assured. When the continuum
limit is approached 
$\| 1-U_{\mu\nu}(x)\|\rightarrow 0$. 
Thus, if we chose
to restrict each plaquette variable by equation 
(\ref{plaqbnd}) the continuum limit would remain
intact.

\subsection{Global Anomalies in 4D}
In the continuum the space of gauge fields
is contractible, but the space of gauge orbits
may not be so. 
An important example, due to Witten \cite{su2anom}, occurs
for $SU(2)$ and has far reaching consequences 
when we have a single Weyl fermion
in the representation $I=1/2$. The space of
gauge transformations is disconnected and
this makes the space of gauge orbits multiply
connected. Although the chiral determinant is
real, it has a twist and gauge invariance cannot
be maintained. The existence of the twist can be
established by using a spectral flow argument
to show that every section will have a 
net zero of degree one on some gauge orbit. 

On the lattice the space of gauge transformations
is connected, but the space of orbits still can
be multiply connected 
because one needs to excise gauge configurations
for which the spectral 
gap at zero in $H$ closes. For $I=1/2$ 
there is a global basis choice for which
the matrix $H$ is real \cite{su2real}, and hence
the overlap has no local phase ambiguity. 
Globally however, the
overlap again defines a twisted bundle (this
time it is a $Z(2)$ bundle) and Witten's anomaly
is reproduced as the sign switch associated with
conic degeneracies in real hamiltonians \cite{ovsu2}. This
generic type of 
sign switch has been discovered by Herzberg and
Longuet-Higgins \cite{conedeg}.

\subsection{$4D$ Noncompact Chiral $U(1)$}

The first four dimensional anomaly ever 
understood was for $U(1)$.
This is a perturbative anomaly and in perturbation
theory it does not matter whether the $U(1)$ is
compact or not. The much more recent
topological understanding of anomalies does not really
apply to noncompact $U(1)$ on infinite $R^4$. 
The $U(1)$ anomaly can be understood in terms of
local cohomology. By starting from some good
regularization scheme one finds a violation 
of gauge invariance in the phase of the chiral determinant.
The question then arises whether one could 
add to the original action a local
functional of the gauge field so that its 
gauge variation cancel the anomaly. It turns
out that the anomaly is inevitable because it cannot be obtained from
the gauge variation of any local functional. 
The anomaly
itself is a local gauge invariant 
functional of the gauge field,
proportional to $\epsilon_{\mu\nu\rho\sigma} 
F_{\mu\nu}F_{\rho\sigma}$. $F$ denotes the field strength, which
is a two form, $F=dA$, where the gauge potential $A$ is
a one form. The anomaly, $(dA)^2$ obviously is closed,
meaning $d(F^2 )=0$. 
If there were a local, gauge invariant,
three form $\omega$ such that $F^2=d\omega$ 
our counterterm would be 
$\int A\omega\propto\int
\epsilon_{\mu\nu\rho\sigma}A_\mu \omega_{\nu\rho\sigma}$.
There seems to be nothing wrong with the equation
$F^2=d\omega$: it is consistent both with $d(F^2)=0$ and
with $\omega$ being gauge invariant. One is free
to choose the four unknown components of $\omega$ 
to reproduce the single component $F^2$. It is obvious that
one can find some solutions, however, all will be
nonlocal functionals of the field strength. One can
classify candidates for anomaly functionals like
$F^2$ by looking for local, gauge invariant, 
functionals of the gauge field that are 
closed but not exact. They define the local 
cohomology of the system. 
In four dimensions there is only one
bad functional, $F^2$. So, if we can 
eliminate it, any other ``anomaly'' will be removable. 

For noncompact chiral $U(1)$ gauge theory the above
logic also works on the lattice \cite{luscher}. If the charges of
all Weyl fermions do not satisfy the well known continuum
anomaly cancelation condition $\sum q^3 =0$ one can
convince oneself that gauge invariance cannot be restored.

A key role in the argument is played 
by another geometrical object, the Berry connection
associated with the $\vert v \rangle$ bundle
over gauge fields, ${\cal A}=i\langle v | dv\rangle$ \cite{geom}. 
This connection has a field strength, Berry's curvature,
and the latter is gauge invariant because it is
independent of the phase of $\vert v \rangle$. Viewing
the space of gauge fields as the base manifold, 
${\cal F}=d{\cal A}$ is obviously exact. The
curvature is defined also over orbit space, but there,
while still closed, it is not necessarily exact if 
space time is compact. In the case when $\sum q^3=0$
the total ${\cal F}$ should be 
exact, since it would be such
if ${\cal A}$ were gauge invariant, and this would be
true if a local 
gauge invariant phase choice were possible. 
If anomalies do not cancel it has been shown that there
are noncontractible closed two dimensional manifolds in
the space of gauge orbits over which the integral of
${\cal F}$ is a non-zero integer. 
This effect takes place only when we work 
on a finite torus.
It proves that in the case of the torus 
a gauge invariant phase choice is impossible if anomalies
do not cancel. 

However, if the anomaly cancelation 
condition is met, a gauge invariant 
choice for the product of chiral determinants associated
with the individual fermions is possible.
The construction is quite technical and works
completely only on an infinite lattice. 
One needs to start from an initial phase 
choice which
gives an anomaly for each fermion. These anomalies do
not cancel even if $\sum q^3 =0$. The initial phase
choice best suited for this problem is an adiabatic 
phase choice \cite{u1neub}. It produces relatively simple formulae
for the anomalies. One can show then that the condition
$\sum q^3 =0$ eliminates the one irremovable lattice
version of $F^2$, and the rest of the total anomaly
can be canceled by a local functional, which enters
as an improvement over the original phase choice. 
Note the similarity of this procedure to the
situation in odd dimensions. 

In four dimensions one faces an additional complication
in the $U(1)$ case \cite{u1neub}. 
The theory does not have a
selfconsistent interacting continuum limit due to
ultraviolet ``triviality''. In the more interesting
non-abelian case this problem would not arise, but also
the simplification of a noncompact version would
be unavailable. For $U(1)$, whether anomalies cancel
or not, the theory only exists (even on paper) as an
effective theory, with a physical scale of energies
$E_{\rm ph}$ and a cutoff $\Lambda$. Triviality means
that the physical coupling constant $e^{\rm ph}$ has
to vanish as $\Lambda$ is taken to infinity. The
distinction between the gauge invariant and gauge 
noninvariant cases now is quantitative. If anomalies
do not cancel we have
\begin{equation}
e^{\rm ph} \le c_1 \left ( {{E^{\rm ph}}\over \Lambda}
\right )^{1\over 3},
\end{equation}
but if they do, we have a much weaker restriction:
\begin{equation}
e^{\rm ph} \le { {c_2} \over 
{ \log\left ( {{\Lambda}\over {E^{\rm ph}}}
\right )}}.
\end{equation}

\subsection{Majorana-Weyl Fermions}

In two and ten dimensional Minkowski space  
one can impose on Dirac fermions simultaneously
a Majorana and a Weyl condition, and be left 
with propagating fermionic degrees of
freedom. The Euclidean version of 
theories containing
Majorana-Weyl fermions in  interaction with gauge fields
can also be regulated on the lattice using the overlap. 
In the appropriate
dimensions, if one starts with Weyl fermions in a real
representation of the gauge group, one can factorize
the overlap into two factors, each representing a single
Majorana-Weyl fermion. This certainly is important
for ten dimensional 
supersymmetric models, but
to keep things simple, 
we shall discuss two dimensions.

Starting from a real set of link matrices
one can rewrite ${\hat H}$ as
\begin{equation}
\hat H = {1\over 2} \hat\alpha^\dagger H \hat\alpha
+{1\over 2} \hat\beta^\dagger H \hat\beta ,
\end{equation}
where the spinorial components of 
$\hat\alpha,\hat\beta$
no longer are independent, 
but are one the hermitian conjugate of the 
other. A similar representation exists 
for $\hat H^\prime$. The total Fock space naturally
factorizes, and so does the overlap. Keeping
one of the factors it is convenient to introduce
hermitian operators, $\hat\gamma$, replacing 
$\hat\alpha$ and 
$\hat\alpha^\dagger$, and generating
a Clifford algebra.  

The overlap description of the Weyl-Majorana \cite{weymaj}
case is based on the many-body hamiltonian
\[
\hat H = {1\over 2} \hat\gamma H \hat\gamma,~~~
H=\Gamma_3 D_W,~~~D_W=m+\sum_\mu (1-V_\mu ),~~~\Gamma_3 =-\sigma_2,
\]
\begin{equation}
V_\mu = V_\mu^* =(V_\mu^T)^{-1}= {{1-\Gamma_\mu}\over 2}
V_\mu + {{1+\Gamma_\mu}\over 2}V_\mu^T,~~~
\Gamma_1=\sigma_3,~\Gamma_2=\sigma_1 .
\end{equation}
with $-2<m<0$ and $H^\prime = \Gamma_3$. 
$H$ and $H^\prime$ are 
hermitian and antisymmetric. By conjugation 
with a real orthogonal matrix $O$, $H$ 
can be brought to an elementary block 
diagonal form with $\sigma_2$ matrices times 
non-negative numbers strung along
the diagonal. $H^\prime$ is already in this form. 
  
Multiplying all 
$\hat\gamma$ operators one obtains
a ``parity'' 
operator $\hat\gamma_{\infty+1}$.
$\hat\gamma_{\infty+1}$ squares to unity and commutes
with $\hat H$ and $\hat H^\prime$. 
If $\det O=-1$, $\vert v\rangle$
and $\vert v^\prime \rangle$ 
are eigenvectors of $\hat\gamma_{\infty+1}$
of opposite sign and
the overlap vanishes, reflecting a mod(2) 
continuum index theorem. 

Once again we see that the overlap is able to 
accommodate rather subtle continuum features.
The mod(2) index theorem means 
that an infinite antisymmetric matrix can
still be meaningfully odd or even dimensional,
and which it is can depend on the gauge field
background. No approach based on a finite number
of fermions per site could reproduce this.

\section{TWO DIMENSIONAL ABELIAN MODELS}

Theoretically, the overlap construction has overcome
major obstacles and dispelled the folklore that
chiral symmetry cannot be realized on the lattice.
But, there still remains the question whether this
progress can be put to use in numerical simulations
that would be practicable at least 
in the foreseeable future. In addition, in the chiral
context, some issues of principle regarding the
effects of gauge violating phase choices can be
tested only by computer simulation. 

One should start gaining numerical experience
from the simplest models that have chiral 
symmetry. This points us to models with chiral
fermions in two dimensions.

\subsection{Vector-like Case}

One of the most important results of the
interplay between chirality and gauge fields
is the explanation why the $\eta^\prime$
cannot be thought of as an approximate Goldstone
boson, while the $\eta$ can. A similar effect
occurs in the massless Schwinger model in two
dimensions, where an apparent global chiral $U(1)$
symmetry is invalidated by fluctuations in the
topology of the gauge field background. This
also happens in a generalization of the model
to several flavors.  With one flavor $\bar\psi\psi$
acquires an expectation value, also at finite Euclidean
volumes - for more flavors, $N_f$, the 
condensate consists
of $N_f$ factors of $\psi$ and $N_f$ 
factors of $\bar\psi$. These condensates can be 
exactly calculated in the continuum for any size and
shape of the 
Euclidean two-torus and they are a direct
consequence of instantons. 

On a finite lattice this effect cannot be
recovered from a quenched simulations, where
fermion loops are ignored. 
(Actually, the quenched
model is theoretically more subtle than 
the original one \cite{quenchschw}.) In the course of a Monte-Carlo
computation one can identify which 
gauge field configurations are responsible for 
the condensate. Thus, we see the Atiyah-Singer 
theorem at work on the lattice, in a dynamical context.

The simulation produced results entirely consistent
with the continuum calculation and with quite high
accuracy \cite{schwvec}, so that the agreement could not be fortuitous.
The parameters of the simulation were chosen so that
one could simultaneously get close enough to continuum
and still keep the fluctuation in the overlap representing
the fermion determinant sufficiently small to permit
the inclusion of the determinant in the observable. 
The fluctuations of the determinant were 
further diminished by exploiting continuum exact results
which made it possible to multiply the 
determinant entering the observable by a certain
function of the gauge background and correcting the action
accordingly. 

The success of this first simulation showed 
that quantum fluctuations in the gauge fields
are not so violent as to erase topological
effects expected from the 
continuum, semiclassical analysis.

The Schwinger model has been studied more recently \cite{rebbisch} 
using a simpler representation of the overlap which
became available in the meantime.

\subsection{Chiral Case}

In the chiral version of the Schwinger model one
has several right handed and several left handed
Weyl fermions interacting through a $U(1)$-gauge field.
Perturbative anomalies would cancel if $\sum_R q_R^2=
\sum_L q_L^2$. If we were working on infinite Euclidean
space and the $U(1)$ gauge field action were of the
non-compact type, starting from an adiabatic phase choice,
we could pick an improved phase convention so that
the total chiral determinant is gauge invariant.

But, to be practical on the lattice, we need to work
on a finite toroidal lattice. Also, it is natural to
make the gauge dynamics that of compact $U(1)$. This
induces some complications. The adiabatic phase choice
is no longer universally applicable, 
because it would connect gauge field
sectors of different topology by going through gauge
field configurations that have $\det(H)=0$. One
needs to resort to some other gauge choice; we chose
a Wigner-Brillouin phase convention \cite{ovnpb2}:
\begin{equation}
\langle v(U\equiv 1) \vert e^{\hat 
v_{\rm thooft} } \vert v(U)\rangle > 0 .
\end{equation}
Here, $\hat v_{\rm thooft}$ is a 't Hooft local vertex
operator summed over the lattice. $\hat v_{\rm thooft}$
is the lowest dimension local 
Lorentz scalar made out of
$\hat a$ and $\hat a^\dagger$ carrying the minimal
amount of $U(1)$ charge that can be violated in an 
elementary topology induced process. Under a gauge
transformation $U\rightarrow U^g$, which acts on 
$H$ by $H(U)= G(g) H(U^g ) G^\dagger (g)$,
and is implemented in Fock space by $\hat G(g)$, 
\begin{equation}
\vert v(U)\rangle = e^{iS_{WZ} (U,g)} . 
\hat G(g) \vert v(U^g )\rangle
\end{equation}
The phase prefactor is the lattice expression of the 
Wess-Zumino functional with the Brillouin-Wigner phase
choice because
\begin{equation}
\langle v^\prime \vert v(U)\rangle =
e^{iS_{WZ} (U,g)}\langle v^\prime \vert v(U^g )\rangle .
\end{equation}
The Brillouin-Wigner phase choice was chosen because
it can be shown analytically to preserve many discrete
symmetries, restricting $S_{WZ}$ in a manner consistent
with continuum, and because it can be 
implemented numerically in a straightforward 
manner \cite{ovnpb2}. The main 
disadvantage of this phase choice is that the definition
does not hold on all gauge field configurations; 
however, the subset on which it fails to hold has zero measure
in the path integral. 
The Brillouin-Wigner phase choice is also
well suited to perturbative calculations. 

Even if perturbative anomalies cancel, the sum of all $S_{WZ}$
contributions on the lattice will generically not. 
So long the gauge action is compact, 
gauge invariance can be restored on the lattice  
by gauge averaging.
In a good theory, gauge invariance could have been
restored even without gauge averaging, by adding
to the phase of the product of all overlaps a certain
local functional of the link variables. It is 
difficult to construct the required functional, but
if it is not too large, gauge averaging would only
produce answers that differ by terms that vanish
in the continuum limit, by a mechanism due to 
F{\" o}rster Nielsen and Ninomiya \cite{fnn}.

The toroidal structure of space-time in conjunction
with the compactness of the gauge group induces some
complications in the chiral case. These complications
seem to imply that there are two dimensional chiral
models that, although consistent at infinite Euclidean
volume, nevertheless are not entirely consistent on 
a compact torus. 
Luckily, there are chiral models that are free of this
difficulty, and a good example is the 11112 model \cite{chiu1npb}.

On a torus there are two angular degrees of freedom
associated with elementary ``Polyakov loops''. These
degrees of freedom can be turned on in a background
that has no field strength and can be traded for
twists in the boundary conditions for the fermions. 
Even though there is no background field strength,
anomalies still need to cancel in the continuum to 
make the total chiral determinant gauge invariant. 
The overlap with the 
Brillouin-Wigner 
phase choice was shown to reproduce the correct continuum
$\theta$-function structure of the chiral determinant \cite{twists}.
Studies of gauge invariance 
restoration showed that the cases
with problems in the continuum behaved badly on the
lattice and ought to be avoided. On the other hand,
the 11112 model for example, works fine. Thus,
this model was selected for a full dynamical simulation.

\subsection{Massless Composite Fermions}

The 11112 model exhibits one of the most 
interesting applications of chirality: when chiral 
symmetries are combined with
confining forces one often ends up 
producing massless fermions that are
bound states of other massless fermions. 
In addition, in the continuum the
11112 model has a combined chiral determinant
that is positive. On the lattice, before
gauge averaging, the product of overlaps can
have a phase, and to again include the chiral
determinant in the observable it is important
that also the fluctuations in the phase
be small. This can be arranged for because the
phase vanishes in the continuum limit. 

The continuum action of the 11112 model 
is given by
\begin{equation}
S={1\over{4e_0^2}} \int d^2 x F_{\mu\nu}^2
-\sum_{k=1}^4 \int d^2 x \bar\chi_k\sigma_\mu
(\partial_\mu +iA_\mu )\chi_k -
\int d^2 x \bar\psi \sigma_\mu^* (\partial_\mu +
2iA_\mu )\psi, 
\end{equation}
where $\sigma_1=1,~\sigma_2=i$. 
Massless neutral bound states are created
by $\bar \eta_{ij}=\bar\chi_i\bar\chi_j\psi$
and $\eta_{ij}=\chi_i\chi_j\bar\psi$ and each
provide a six dimensional representation of the
global $SU(4)$ acting on the $\chi$-fields. 
At first one would guess that the $\bar\eta$ and
$\eta$ fields create massless left-handed
Weyl fermions. But, the $SU(4)$ anomaly carried
by the left-handed Weyl $\chi$-particles would
not match unless the $\eta$-particles are 
left-handed Majorana Weyl rather than just Weyl.
This requires additional mixing between the
$\eta$ particles, possible only if fermion number
is not conserved. Instantons are present in the
models and they indeed induce the necessary
violation and mixing. The instanton effect
makes the 't Hooft vertex,
\begin{equation}
v={{\pi^2}\over{e_0^4}} \chi_1\chi_2\chi_3\chi_4
\bar\psi (\sigma\cdot\partial )\bar\psi 
\end{equation}
and its conjugate get an expectation value. In
terms of $\eta$, $v$ looks like an 
off-diagonal kinetic energy term, 
$\epsilon_{ijkl} \eta_{ij}(\sigma\cdot\partial )\eta_{kl}$. 
The massless Majorana Weyl 
composites are $\rho_{ij} \propto [\eta_{ij}
-{1\over 2} \epsilon_{ijkl}\bar\eta_{kl} ]$; the
other linear combination has no massless poles
in its propagator \cite{compplb1}. 

The expectation value of the 't Hooft vertex
operator can be calculated in the continuum both
at infinite volume and on a torus \cite{compplb2}. The continuum
number has been obtained on the lattice using 
the Brillouin-Wigner phase definition and gauge
averaging in a full dynamical simulation \cite{compprd}.

\section{QCD}

Lattice techniques are unique in that they provide, in principle,
a method of calculating to any accuracy various properties of
a strongly interacting field theory like QCD. This is important
not only in itself, but, QCD effects mask almost all minimal
standard model predictions and need to be ``pealed off'' before
one can extract from experiment the numerical value of quark
masses and mixings. Steady progress over the last twenty years
has made numerical QCD into a competitive quantitative method
in estimating strong interaction effects. Among other problems,
numerical QCD also suffered from the difficulty  to make global
chiral symmetries only weakly broken on the lattice, just as it
is in the continuum. This problem has been solved in principle
and the solution is already marginally practicable. Further
advances, including the certain increase in computational power,
will eventually make it possible to incorporate the new lattice
chirality into most numerical QCD projects.

\subsection{Instantons}

One of the first areas where the new formalism 
has an advantage is in identifying instanton effects on
the lattice. In QCD instanton effects are important mainly
because of their impact on fermions and the key ingredient
is chirality. In the background of an instanton the Weyl-Dirac
operator governing the dynamics of, say, the right handed
component of a massless quark, should be thought of as an
infinite matrix with one more row than columns. This situation
is represented by having $N_v$, the dimension of the negative
eigenspace of $H$, differ from $N/2$ by one unit. This happens
when the mass parameter in $H$, $m$, is around $-1$. For
positive $m$, $N_v=N/2$ always. Therefore, as $m$ is decreased
from a slightly positive value to about $-1$, the lowest
positive eigenstate of $H$ crosses zero and becomes negative.
The same is true if the background is more general, but still
carries the same amount of topological charge as an instanton would.

The flow of eigenvalues as a function of the mass parameter $m$
provides a relatively inexpensive method for numerically finding
the topological charge, $n_{\rm top}$, in the overlap definition.
Several studies in two and four dimensions have been made using this
method and the results were quite useful 
\cite{ovnpb2,prl1,vranar}. In principle, the main
advantage of the method is that it uses a definition of topological
charge on the lattice that is entirely consistent with the response of 
the overlap lattice fermions. Thus, even if the gauge background is
quite ``rough'', and a precise association with any topological
number somewhat questionable,  there is no question that precisely
this background contributes to the $\eta$-$\eta^\prime$ mass difference,
on the lattice, before any continuum limit is taken.

Superficially, the flow method is similar to a 
method used before \cite{japflow}
to estimate topological charge. There, one used traditional Wilson
fermions with a mass $m$ tuned to the vicinity of its critical value.
Instantons were related to approximate zero modes of the hermitian
Wilson-Dirac operator with substantial chirality. A fermion state was
considered an approximate zero mode if it was the eigenstate of
the hermitian Wilson Dirac operator that would cross zero for some
$m$ numerically close to the critical $m$ and had substantial chirality.
In the overlap we look at the same eigenvalue flows, but not only in
the vicinity of the critical $m$.

\subsection{The Overlap Dirac Operator}

There were two essential reasons against having a traditional
lattice action for Weyl fermions. The first was that the global
$U(1)$ associated with the Weyl fermions could be gauged and no
room for anomaly was left. The second was that an instanton background
changed the effective shape of the Weyl-Dirac operator from square
to rectangular. In a vector like theory we have a pair of conjugate Weyl
fermions. There still exists an objection against a traditional action,
but it is weaker now: One should still be prohibited
from gauging the axial $U(1)$. On the other hand, the pair of
Weyl fermions are controlled by Weyl-Dirac 
matrices of complementary rectangular
shapes, so that, when packed together to make the Dirac operator, 
they produce a square matrix of fixed size. 
From the overlap we know that the global $U(1)$ is avoided
since the fermionic number of $\vert v\rangle$ can be unequal to that
of $\vert v^\prime\rangle$. This means that the global chiral $U(1)$ is
not quite realized. We conclude that it should be possible to have
a more traditional formulation in the vector-like case than in the chiral
case. All that has to happen is that the global $U(1)$ must not quite
hold, but appear in a way realized by the overlap. So long there is no
obvious way to gauge this $U(1)$ we should be fine. 

Looking for a more traditional action for the vector-like case we
observe that in one dimension higher the overlap formulation
simplifies as we have seen and one has a simple natural formula for
the five dimensional kernel $v^{\prime\dagger}v$. 
Dimensionally reducing this formula
produces a field with the right number of components to be a Dirac
field in even dimensions. We therefore are led trivially to the
following action (for definiteness in four dimensions) \cite{ovlapdir}:
\begin{equation}
\bar\psi D_o\psi,~~~D_o={{1+V}\over 2},~~ 
V=D_W{1\over \sqrt{D_W^\dagger
D_W}}=\gamma_5 {\rm sign }(H) .
\end{equation}
Actually, recalling that the five dimensional case was derived from the
six dimensional overlap, we can dimensionally reduce by 2 from
six dimensions directly and obtain the same formula. A Weyl fermion
in six dimension has the same number of components as a Dirac fermion
in four.  
The overlap Dirac operator $D_o$ does not anticommute with $\gamma_5$,
but clearly represents massless fermions because it comes from Weyl
fermions in six dimensions.  $D_o$ does have a special relation
with $\gamma_5$: The propagator $D_o^{-1}$ anticommutes with $\gamma_5$
up to a local contact term, as a result of $\gamma_5 V\gamma_5=V^\dagger$. 
While the contact term has no influence
on low energy consequences of chirality, it prohibits the construction
of an exactly axial-$U(1)$ gauge invariant theory. 
\begin{equation}
\{{D_o}^{-1} ,\gamma_5 \}=
\gamma_5 [{2\over {1+V^\dagger }} +{2\over{1+V}}]=2\gamma_5 ,~~~
\{{D_o}^{-1} -1  ,\gamma_5 \}=0 .
\end{equation}
Unexpectedly, the above relations turned out \cite{vectorprd, moreov}
to be a particular
case of a proposal by Ginsparg and 
Wilson made already in 1982 \cite{ginsparg}. This time
there is no doubt that the desired $D_o$ exists, so long $H$ is gapped
around zero. More recently, it was suggested that another
action satisfying the Ginsparg Wilson criterion is provided
by a fermionic ``perfect'' action \cite{hasen}. This action
is allegedly obtainable as a limit of an iteration process that
acts also on the gauge field background. There is no explicit formula,
and the iteration process involves a non-analytic step of maximization.
It is unclear that the iteration process has a limit, that such
a limit, if it exists, is unique
and it is worrisome that analyticity in the gauge field background is 
sacrificed in each iteration. Moreover, the precise set of requirements,
beyond that of satisfying the Ginsparg Wilson relation, a fermionic action
needs to obey in order to really represent one massless Dirac fermion
on the lattice is not fully understood \cite{twaichiu}. There are known
actions that do satisfy the Ginsparg Wilson relation but do not represent
massless Dirac fermions \cite{moreov}. 

With the help of $D_o$ the formula $n_{\rm top} = -{1\over 2} Tr \epsilon$
\cite{ovnpb1,ovnpb2,prl1} can be rewritten as $n_{\rm top} =
-Tr \gamma_5 D_o$ and the $Tr$ operation can be rewritten as 
$Tr O = \sum_x tr O_{xx}$ where $tr$ denotes a trace over spinorial
and group indices and $x$ is a lattice site. 
The quantity $\gamma_5 {D_o}_{xx}$ is the divergence of the
covariant current defined in \cite{geom}. The manipulations
required to see this explicitly can be found in \cite{dubna}.

\subsection{Truncation I: Domain Walls}

While $D_o$ looks simple and is a concrete resolution to the
chirality problem in the vector-like context, its practical
usefulness seems unclear: the sign function of $H$ is not easy
to compute numerically, and it is not a sparse matrix. Somehow,
this exact expression has to be truncated to make it practical
for use. Since we got to the overlap by making the fifth direction
in Kaplan's setup (modified by taking the gauge fields to be truly four
dimensional) infinite the first thing to analyze is the case when
we make the extent of the fifth dimension finite again. 
As is clear from the original paper by Kaplan
and from other work, one gets a system containing many Dirac
fermions, with one of them very light. The mass of the light fermion
goes to zero exponentially with the length of the fifth dimension.

A more precise description of the domain wall setup is that it
describes a finite, but large, number of Dirac flavors, whose
action is local and represented by a sparse matrix, but who are
mixed by a gauge field dependent mass matrix in such 
a way that when the number of
flavors goes to infinity, the mass of one of the Dirac fermions
goes to zero in a way that requires no fine tuning, while all
the rest of the Dirac fermions (including doublers) 
keep a mass of the order of the
inverse lattice spacing. This must mean that the effective
action for the light fermion is a truncation of some version
of the overlap Dirac operator, a truncation that becomes exact
in the limit of an infinite number of flavors. It is important
to understand the mechanism that ensures masslessness in the
limit, without fine tuning.  To this end we need some notation:
\begin{equation}
\gamma_\mu =\pmatrix{0&\sigma_\mu\cr\sigma_\mu^\dagger &0\cr},
~~~C=\sum_\mu \sigma_\mu (T_\mu - T_\mu^\dagger ),~~~
B=m+\sum_\mu (1-{{T_\mu+T_\mu^\dagger}\over 2}) .
\end{equation}
The number of flavors is $k$. The matrices $B$, $C$ are $q\times q$
where, for $SU(3)$, $q=6V_l$, with $V_l$ being the total number of sites
in the four-dimensional space-time lattice. The action is
\begin{equation}
S=\bar\psi D(0,0) \psi ,
\end{equation}
where the matrix $D(0,0)$ is a special case of $D(X,Y)$ with $X$ and $Y$
general $q\times q$ matrices. 
\begin{equation}
D(X,Y)=\pmatrix{ C^\dagger & B & 0 & 0 & 0 & 0 &\ldots &\ldots & 0 & X\cr
            B 	     & -C & -1& 0 & 0 & 0 &\ldots &\ldots & 0 & 0\cr
	    0   &-1& C^\dagger& B & 0 & 0 &\ldots &\ldots & 0 & 0\cr
            \vdots&\vdots&\vdots&\vdots&\vdots&\vdots&\ddots&\ddots
&\vdots&\vdots\cr
	    Y & 0 &  0 & 0& 0         & 0 &\ldots &\ldots & B &-C\cr}.
\end{equation}

A key formula is 
\begin{equation}
\det D(X,Y) = (-)^{q(k-1)} (\det B)^k \det 
\left [ \pmatrix {-X &0 \cr 0& 1 } - T^{-k} \pmatrix {1&0\cr 0 &-Y\cr}
\right ]
\end{equation}
with the transfer matrix $T$ given by 
\begin{equation}
T=e^{-H}=\pmatrix {{1\over B} & {1\over B} C \cr
         C^\dagger {1\over B} & C^\dagger {1\over B} C +B \cr} .
\end{equation}
The determinant of $D(0,0)$ contains
contributions from the light fermion and from all heavy ones. Setting
$X=Y=\mu$ with $\mu <<1$ gives the light fermion a mass of order
$\mu$ at $k=\infty$. Setting $X=Y=1$ makes all fermions heavy. Thus,
one can associate with the massless fermion the ratio \cite{vectorprd}
\begin{equation}
{{\det D(0,0)}\over{\det D(1,1)}} = \det \left [ {1\over 2}
\left ( 1+\gamma_5 \tanh ({1\over 2} k H ) \right )\right ] .
\end{equation}
The division by $D(1,1)$ can be viewed as the consequence of
additional $k$ Dirac fields of wrong statistics. 

So long $H$ has no zero eigenstates the $k=\infty$ limit is seen
to be given by the determinant of an overlap-type operator, only
the form of $H$ is different (here $H=-\log (T)$). 
The difference is irrelevant to the
continuum limit. By varying with respect to $X$ and $Y$ one can obtain
light fermion propagators. Setting $X=Y=\mu$ and differentiating
produces a formula for the physical condensate. 
Repeating the construction
for several physical flavors one gets:
\begin{equation}
{1\over {N_f}}
\sum_{f=1}^{N_f} < (\bar\psi_f \psi_f )_{\rm physical} > =
{{\cal Z}\over{ L^d}} <{\det}^{N_f} 
\left [ {{1+V}\over 2}\right ] {\rm Tr} {{1-V}\over{1+V}}
>_{U} .
\end{equation}
Here $<..>_{U}$ means an average with respect to the pure gauge action
and ${\cal Z}$ is a renormalization constant. An instanton produces
an eigenstate of $V$ with eigenvalue $-1$. 
We see how single
instantons would give a nonzero contribution for one flavor,
but no contribution for more flavors, just as expected from
the more formal continuum expressions. If topology is trivial,
conjugation by $\gamma_5$ produces a sign switch,
as a consequence of $\gamma_5 {{1-V}\over{1+V}} \gamma_5 =
-{{1-V}\over{1+V}}$. Connection to the Ginsparg Wilson relation
is made through ${{1-V}\over{1+V}}={2\over{1+V}}-1$. 

This analysis easily generalizes to the supersymmetric case where 
the gauge link variables can be chosen real. Global choices
of basis exist then that make the $D$ matrices antisymmetric
and, roughly, the determinants get replaced by pfaffians \cite{ovnpb2,
vectorprd}. 

The mechanism assuring masslessness without fine tuning is a generalized
see-saw \cite{vectorprd}. 
Making some further basis changes, the mass matrix can be
brought to a hermitian form:
\begin{equation}
M=\pmatrix { 0&0&\ldots&0&0&B\cr
	     0&0&\ldots&0&B&-1\cr
	     \vdots&\vdots&\vdots&\vdots&\vdots&\vdots\cr
	     B&-1&\ldots&0&0&0\cr} .
\end{equation}
In the absence of gauge fields, $B$ is small for momenta near zero
but exceeds unity in the vicinity of the doublers. When $|B|$ is
smaller than unity $M$ has one eigenvalue of order $|B|^k$ and all
other eigenvalues of order 1. One does not need to fine tune; so long
$|B|$ is smaller than one and $k$ is large enough a very light
Dirac fermion is present, at least for weak gauge fields. It is
not entirely trivial, in this way of looking at things, why
the mechanism is stable under radiative corrections. To one loop
this has been checked, and at most the radiative correction changes
the light fermion mass from $\sim e^{-k ~{\rm Const}_1}$
to $\sim k^{{\rm Const}_2} e^{-k ~{\rm Const}_1}$, where the constants are
positive \cite{kikukawa}. 

The main advantage of using domain walls is that one has a local action
of traditional form and tested numerical methods can be applied.
Theoretically,
the domain wall setup is appealing because the truncation has a 
physical interpretation: the
two walls on which the two Weyl components reside 
are separated by a finite rather than infinite distance. 
The approximation is limited by how small the
eigenvalues of $H^2$ can get in typical gauge fields. Perturbative
calculations at finite $k$ are tedious. One cannot focus on the light
fermion with ease because the structure of $H$ is awkward.

\subsection{Truncation II: Overlap}

It makes sense to look for other truncations, which are closer
to the overlap with the $H$ given by the hermitian Wilson-Dirac
matrix. There is no reason to seek a physical interpretation of
the truncation, like in the domain wall context. We are simply
faced with a problem in numerical analysis: how do we 
calculate to machine accuracy 
the action of ${\rm sign} (H)$ on an arbitrary vector
when $H$ is sparse and ${\rm sign} (H)$ is too large to fit in
the computer's memory. The first observation is that achieving
machine accuracy will be very expensive if $H$ is ill conditioned.
Only the condition number enters because obviously the sign function
is invariant under a positive rescaling of $H$. The most direct way
is to look at the sign function as any other special function that
we need to implement on the machine. This suggests a rational 
approximation. Since we have to deal with a possibly ill conditioned
matrix, we would like to exploit one of the better tested numerical
methods, which is the conjugate gradient algorithm (CG). Thus,
we look for a rational approximation to the sign function that
can be written as a simple sum of pole terms. Such an approximation
can be obtained starting from a Newton iteration algorithm for
the sign function \cite{prl2,higham}: 
\begin{equation}
\epsilon_n (z) 
={{(1+z)^{2n} -(1-z)^{2n} }\over { (1+z)^{2n} + (1-z)^{2n}}}=
{z\over n} \sum_{s=1}^n {1\over{
z^2 \cos^2 {\pi\over {2n}} (s-{1\over 2}) +\sin^2 {\pi\over {2n}} 
(s-{1\over 2})}} .
\end{equation}
Clearly, so long $z\ne 0$, $\epsilon_n (z)$ approaches the sign
of $z$. The approach is exponential in $n$, but slows when $|z|$
is very small or very large. 
All needed inversions would involve $H^2$ 
shifted by varying positive constants. These inversions
can all be done simultaneously at a total computational cost equal
to that of the slowest inversion. Only memory usage grows linearly
with $k$. $k$ also controls numerical accuracy. 

The linear growth of memory may be a problem because even if main
memory suffices, the faster accessible memory in caches is always
severely limited. One can trade memory usage for operations. After
all, the CG is a numerically stable application of the Lanczos 
algorithm and it is known that in the latter one can resort to
a ``two pass version'' in situations where a single pass approach
would have a large memory cost. Thus, at the cost of about a factor
of two in operations, also the memory demands are limited and $k$
independent. The difficulty of the problem has now been reduced to
its essential part, namely the condition number of $H$ \cite{twopass}.

If one applies the above method in a dynamical fermion simulation
one needs to compute the inverse of $D_o$ very often. This results
in a two level nested CG procedure, the outer CG computing the
inverse of $D_o^\dagger D_o$ and the inner the sign of $H$. This seems
cumbersome when compared to domain wall fermions, where a single
CG is needed, albeit on a sparse matrix which is of order $k$ larger
than $H$. Whether there is real cost associated with this complication
or not is somewhat unclear. However, even if one is
loath of CG nesting, one can still go about 
implementing the sign function by a rational 
approximation because any such approximation is equivalent 
to an extended system governed by a sparse matrix of larger
size \cite{chain1}. 

For example \cite{chain2}, the system appropriate to the particular representation
used above is:
\begin{equation}
S= \bar \Psi \gamma_5 K \Psi .
\end{equation}
Introducing also a small quark mass $\mu$ and the notation
\begin{equation}
c_s= \cos \theta_s ,~~s_s=\sin \theta_s,~~~\theta_s =
{\pi\over{2n}} (s-{1\over 2}),~s=1,2,...,n ,
\end{equation}
the $K$ matrix can be written as follows:
\begin{equation}
K=\pmatrix{ 
-{{1+\mu}\over 2} \gamma_5 & \sqrt{{1-\mu}\over {2n}}&0&\sqrt{{1-\mu}\over
{2n}}&0&\dots
&\sqrt{{1-\mu}\over {2n}}&0\cr
\sqrt{{1-\mu}\over {2n}}& c_1^2 H & s_1 &0 &0 &\dots &0&0\cr
0& s_1 &-H &0&0 &\dots &0&0\cr
\sqrt{{1-\mu}\over {2n}}&0&0&c_2^2 H & s_2 &\dots& 0& 0\cr
0&0&0&s_2 &-H &\dots&0&0\cr
\vdots& \vdots& \vdots& \vdots& \vdots&\vdots&0&0\cr
\sqrt{{1-\mu}\over {2n}} &0&0&0&0&\dots&c_n^2 H & s_n\cr
0&0&0&0&0&\dots&s_n&-H\cr} .
\end{equation}

Detailed study indicates that the above system is as efficient numerically
as domain walls, but has the advantage that the effective action for
the light degrees of freedom is relatively simple:
\begin{equation}
D_{\rm eff} = {{1+\mu}\over 2} + {{1-\mu}\over 2}\gamma_5\epsilon_n (H) .
\end{equation}
It is also possible that the nested CG versions, with one or two passes
are actually more efficient numerically. Anyhow, the indications are that
domain walls will be eventually replaced in numerical QCD 
by a more direct implementation of the overlap Dirac operator, $D_o$.

\subsection{The Problem of Zero Mode Artifacts}

Any numerical method approximating the sign function of some $H$
will face difficulties when $H$ has very small eigenvalues relative
to 1 (this
presupposes that the largest eigenvalues of $H^2$ are always 
of order unity). Although the $H$ in the overlap and the one
in domain walls are different, their small eigenvalues are closely
related, so the same gauge configurations will be problematic in either
approach. At numerically feasible QCD gauge couplings it is a fact
that there are quite a few states that have very small $H$-eigenvalues 
\cite{scri}.

A first worry about these states is whether they signal that somehow
the lattice has outwitted us again, and has found a way to avoid chirality.
The exact lower bounds on $H^2$ which we discussed before void this
worry. What is left to understand is why these modes occur, how they
affect QCD observables and what can be done to eliminate them from
the numerical process.  

That there are some gauge backgrounds that will produce such modes is
easy to understand as a result of topology. We know that any lattice
gauge configuration can be smoothly deformed to any other so a background
that has non-zero lattice topology $n_{\rm top}$  can be deformed to
unit link matrices everywhere. Necessarily, some eigenvalues of $H$
will have to cross zero during the deformation, producing zero modes
at any $m$ we happened to choose. We also know that taking $m$ from
$0$ to a little less than $-2$ must produce several crossings  if the
background has, say one relatively smooth instanton. This is so because,
say, at $m=-1$ we have one massless Dirac fermion,
so $n_{\rm top}$ will change from zero to one, requiring a zero mode
at some negative $m$ close to zero. Further, after $m=-2$ four more
doublers come into play. Since their frames are reversed in orientation
relatively to the fermion at zero momentum, they see an anti-instanton,
and would produce, by themselves, a $n_{\rm top}=-4$. Taking into account
all massless Dirac particles we should produce a total $n_{\rm top}=-3$
which requires four modes to head in the opposite direction and cross zero
before $m$ reaches, say $m=-3$. Typically, the mode that crossed first
turns around and crosses again, being joined by three more modes.
All in all, there is extensive traffic across zero in the spectrum of $H$
as $m$ is varied throughout the region of interest. This is further 
complicated by the effect of the fluctuations in the gauge fields, which
through the Wilson mass term contribute a stochastic component to the
mass parameter $m$ moving the crossing points around, and effectively
covering the entire range of $m$.  The
exact upper bound for $0<m<-4$, 
$8+m$ \cite{neubound}, 
is saturated in typical gauge configurations. When $m$ decreases,
the spectrum of $H$ is squeezed and level repulsion has a tendency to 
push more states towards zero eigenvalue.

The zero modes are a lattice artifact; their appearance results from
the substantial difference between 
the spaces of lattice and continuum gauge fields. 
The zero modes are well localized and are likely to make only small
contributions to correlations at large distances. Hence, they
ought to have little numerical 
effect on physical observables. Explicit 
examples of localized zero modes and the localized gauge field
configurations that support them are presented in \cite{localzm}.

There are many ways to ameliorate the numerical pain caused by approximate
zero modes to $H$. This is an area of intensive current research.

\section{SUMMARY}

The folklore that chiral symmetries cannot be realized on the lattice
has been almost rigorously proven to be 
false. Chiral symmetry and the lattice get reconciled
by a mechanism that employs an
infinite number of fermions per unit Euclidean four volume. This 
mechanism 
extends beyond lattice regulators, and provides a generic way 
for effective Lagrangians with chiral fermions to appear out of more
fundamental theories which may contain no concept of chirality. 
The solution to the problem of chiral symmetry on the lattice is
almost complete and the remaining open issues
are mostly technical. Perhaps the most physical of the remaining open
problems is to construct a Hamiltonian version of the overlap. 

The key to the breakthrough was to take seriously the indications
from the mathematical insights gained during the mid eighties about
anomalies and about the chiral determinant. 
Crucial was the acceptance of the
viewpoint that the chiral determinant is a section in a line bundle
over orbit space and that the response of fermions to topology indeed
is as semiclassical computations indicated. 

With the benefit of hindsight it appears that one could have solved
the problem of lattice chirality 
much earlier had the suggestion of Ginsparg and
Wilson been scrutinized more thoroughly. Historically, this can be
viewed either as the case of a paper before its time, or assign blame
to the many lattice papers in the decade between the early
eighties to the early nineties that produced failure after
failure, ignored the mathematical progress that was taking
place in the continuum, and ended up reinforcing what now
is widely considered a false folklore. 

\section{ACKNOWLEDGMENTS}

This research was supported in part by the DOE grant DE-FG02-96ER40949.


\begin{thebibliography}{99}
\bibitem{callan} Callan, CG, Harvey, JA, 1985.
Anomalies and Fermion Zero Modes on Strings and Domain Walls.
{\it Nucl. Phys. B250}:427.
\bibitem{kaplan} Kaplan, DB. 1992.
A Method for Simulating Chiral Fermions on the Lattice.
{\it Phys. Lett. B288}:342-7.
\bibitem{slavnov} Frolov, SA,  Slavnov, AA. 1993.
An Invariant Regularization of the Standard Model.
{\it Phys. Lett. B309}:344-50.
\bibitem{ovplb1} Narayanan, R, Neuberger, H. 1993. 
Infinitely Many Regulator Fields for Chiral Fermions. 
{\it Phys. Lett. B302}:62-69.
\bibitem{ginsparg} Ginsparg, PH, Wilson, KG. 1982.
A Remnant of Chiral Symmetry on the Lattice.
{\it Phys. Rev. D25}:2649.
\bibitem{creutzrev} Creutz, M. 2000.
Aspects of Chiral Symmetry on the Lattice.
{\it hep-lat/0007032}. 
\bibitem{contin} Alvarez-Gaume, L, Ginsparg, P.1986.
The Structure of Gauge and Gravitational Anomalies.
{\it Ann. Phys. 163}:228.
Alvarez-Gaume, L. 1985. An Introduction to Anomalies.
{\it Erice School Math Phys}:0093. 
Ball, RD. 1989.
Chiral Gauge Theory.
{\it Phys. Rept. 182}:1.
\bibitem{thooft} 't Hooft, G. 1976.
Symmetry Breaking through Bell-Jackiw Anomalies.
{\it Phys. Rev. Lett.36}:1119.
\bibitem{descent} Stora, R. 1976.
Continuum Gauge Theories.
{\it Cargese Summer Inst.}:0201.
Zumino, B. 1984.
Chiral Anomalies, Higher Dimensions, and Differential Geometry.
{\it Nucl. Phys. B239}:477.
\bibitem{cohom} Brandt, F, Dragon, N, Kreuzer, M. 
All Consistent Yang-Mills Anomalies.
{\it Phys. Lett. B231}:263-270.
\bibitem{nielsen} Nielsen, HB, Ninomiya, M. 1981.
No Go Theorem for Regularizing Chiral Fermions.
{\it Phys. Lett. B105}:219.
\bibitem{eichten} Eichten, E, Preskill, J. 1986.
Chiral Gauge Theories on the Lattice.
{\it  Nucl.Phys.B268}:179.
\bibitem{brs} Alvarez-Gaume, L, Baulieu, L. 1983.
The Two Quantum Symmetries Associated With a Classical Symmetry.
{\it Nucl. Phys. B212}:255.
\bibitem{brsnogo} Neuberger, H. 1986.
Nonperturbative BRS Invariance. {\it Phys. Lett. B175}:69.
Neuberger, H. 1987.
Nonperturbative BRS Invariance and the Gribov Problem. 
{\it Phys. Lett. B183}:337.
\bibitem{ovnpb1} Narayanan, R, Neuberger, H. 1994.
Chiral Determinant as an Overlap of Two Vacua. 
{\it Nucl. Phys. B412}:574-606.
\bibitem{ovnpb2} Narayanan, R, Neuberger, H, 1995.
A Construction of Lattice Chiral Gauge Theories.
{\it Nucl. Phys. B443}:305-385.
\bibitem{boya} Boyanowsky, D, Dagotto, E, Fradkin, E, 1987.
Anomalous Currents, Induced Charge and Bound States on a Domain
Wall of A Semiconductor.
{\it Nucl. Phys. B285}:340. 
Shamir, Y, 1993.
Chiral Fermions from Lattice Boundaries.
{\it Nucl. Phys. B406}:90-106.
\bibitem{prl1} Narayanan, R, Neuberger, H. 1993.
Chiral Fermions on the Lattice. {\it Phys. Rev. Lett. 71}:3251.
\bibitem{daemi} Randjbar-Daemi, S, Strathdee, J. 1995.
On the Overlap Formulation of Chiral Gauge Theory.
{\it Phys. Lett. B348}:543-48. 
Gravitational Lorentz Anomaly from the Overlap Formula in Two Dimensions.
{\it Phys. Rev. D51}:6617-19. 
Chiral Fermions on the Lattice.
{\it Nucl. Phys.B443}:386-416. 
Randjbar-Daemi, S, Strathdee, J. 1996.
On the Overlap Prescription for Lattice Regularization of Chiral Fermions.
{\it Nucl. Phys.B466}:335-360.
Vacuum Polarization and Chiral Lattice Fermions.
{\it Nucl. Phys.B461}:305-326.
Randjbar-Daemi, S, Strathdee, J. 1997.
Consistent and Covariant Anomalies in the Overlap Formulation of 
Chiral Gauge Theories.
{\it Phys.Lett.B402}:134-140.
\bibitem{newpert} Capitani, S, Giusti, L.2001. 
Analysis of the Delta I=1/2 Rule and e'/e with Overlap Fermions.
{\it hep-lat/0011070}.
Perturbative Renormalization of Moments of Quark Momentum, Helicity and
Transversity Distributions with Overlap and Wilson Fermions
{\it hep-lat/0009018}.
Capitani, S, Giusti, L.2000.
Perturbative Renormalization of Weak-Hamiltonian Four-Fermion Operators with
Overlap Fermions {\it Phys. Rev. D62}:114506.
Perturbative Renormalization of the First Two Moments of Non-singlet Quark
Distributions with Overlap Fermions.
{\it Nucl.Phys. B592}:183-202.
Alexandrou, C, Follana, E, H. Panagopoulos, H, Vicari, E.2000.
{\it Nucl.Phys. B58}:394-406.
Lambda-parameter of Lattice QCD with the Overlap-Dirac Operator
{\it Nucl.Phys. B571}:257-266.
Kerler, W. 2000. 
Chiral Fermions on the Lattice and Index relations.
{\it hep-lat/0007023}.
Adams, DH. 2000. Dirac Operator Index and the Topology of Lattice Gauge
Fields, {\it Chi. J. Phys. 38}:633-646.
\bibitem{kikukawa} Kikukawa, Y, Neuberger, H, Yamada, A. 1998.
Exponential Suppression of Radiatively 
Induced Mass in the Truncated Overlap.
{\it Nucl. Phys. B526}:572-96.
\bibitem{aoki} Aoki, S, Izubuchi, T, Kuramashi, Y, Taniguchi, Y. 1999.
Perturbative Renormalization Factors of Three Quark and Four Quark
Operators for Domain Wall QCD.
{\it Phys. Rev. D60}:114504.
\bibitem{mooregaume} Moore, G, Della Pietra, S, Alvarez-Gaume, L.1985.
Anomalies and Odd Dimensions.
{\it Ann. Phys. 163}:288.
\bibitem{kikneu} Kikukawa, Y, Neuberger, H. 1998.
Overlap in Odd Dimensions. {\it Nucl. Phys. B513}:735-57.
\bibitem{neubound} H. Neuberger. 2000. Bounds on the Wilson Dirac Operator.
{\it Phys. Rev. D61}:085015.
\bibitem{su2anom} Witten, E. 1982. 
An SU(2) Anomaly. {\it Phys. Lett. B117}:324-328.
\bibitem{su2real} Neuberger, H. 1998.
Explicitly Real Form of the Wilson-Dirac Matrix for SU(2).
{\it Phys. Lett. B434}:99-102.
\bibitem{ovsu2} Neuberger, H. 1998.
Witten's SU(2) Anomaly on the Lattice.
{\it Phys. Lett. B437}:117-122.
\bibitem{conedeg} Herzberg, G,  Longuet-Higgins, HC. 1963.
Intersection of Potential Energy Surfaces in Polyatomic Molecules.
{\it Disc. Farad. Soc. 35}:77-82. 
\bibitem{luscher} Luscher, M.1999. 
Abelian Chiral Gauge Theories on the Lattice with Exact Gauge Invariance.
{\it Nucl. Phys. B549}:295-334.
\bibitem{geom} Neuberger, H. 1999.
Geometrical Aspects of Chiral Anomalies in the Overlap.
{\it Phys. Rev. D59}:085006.
\bibitem{u1neub} Neuberger, H. 2001.
Noncompact Chiral U(1) Gauge Theories on the Lattice.
{\it Phys. Rev. D63}:014503.
\bibitem{weymaj} Huet, P, Narayanan, R, Neuberger, H.1996.
Overlap Formulation of Majorana--Weyl Fermions.
{\it Phys. Lett. B380}:291-5.
\bibitem{quenchschw} Casher, A, Neuberger, H.1984.
The Density of States of a Two Dimensional Electron Gas in a Random External
Magnetic Field. {\it Phys. Lett. B139}:67. 
Narayanan, R, Kiskis, J. 2000.
Chiral Condensate in the Quenched Schwinger Model.
{\it Phys. Rev. D62}:054501.
\bibitem{schwvec} Narayanan, R, Neuberger, H, Vranas, P. 1995.
A Simulation of the Schwinger Model in the Overlap Formalism.
{\it Phys. Lett. B353}:507-12.
\bibitem{rebbisch} Giusti, L, Hoelbling, C, Rebbi, C. 2000.
Exact Results and Approximation Schemes for the Schwinger Model with the
Overlap Dirac Operator.
{\it hep-lat/0011014}.
\bibitem{fnn} Forster, D, Nielsen , HB, Ninomiya, M. 1980.
Dynamical Stability of Local Gauge Symmetry: Creation of Light From Chaos.
{\it Phys. Lett.B94}:135.
\bibitem{chiu1npb} Narayanan, R, Neuberger, H. 1996.
Anomaly Free U(1) Chiral Gauge Theories on a Two Dimensional Torus.
{\it Nucl. Phys. B477}:521-48.
\bibitem{twists} Narayanan, R, Neuberger, H. 1995.
Two Dimensional Twisted Chiral Fermions on the Lattice.
{\it Phys. Lett. B348}:549-52.
Randjbar-Daemi, S, Fosco, CD. 1995.
Determinant of Twisted Chiral Dirac Operator on the Lattice.
{\it Phys.Lett.B354}:383-8.
\bibitem{compplb1} Narayanan, R, Neuberger, H. 1997.
Massless Composite Fermions in Two Dimensions and the Overlap.
{\it Phys. Lett. B402}320-7.
\bibitem{compplb2} Kikukawa, Y, Narayanan, R, Neuberger, H. 1997.
Finite Size Corrections in Two Dimensional Gauge Theories and a 
Quantitative Chiral Test of the Overlap. {\it Phys. Lett. B399}:104-112.
\bibitem{compprd} Kikukawa, Y, Narayanan, R, Neuberger, H. 1998.
Monte Carlo Evaluation of a Fermion Number Violating Observable in 2D.
{\it Phys. Rev. D57}:1233-41.
\bibitem{vranar} Narayanan, R, Vranas, P,  1997.
A Numerical Test of the Continuum Index Theorem on the Lattice.
{\it Nucl. Phys. B506}:373-386.
\bibitem{japflow} Itoh. S, Iwasaki, Y, Yoshie, T. 1987.
U(1) Problem and Topological Excitations on a Lattice.
{\it Phys. Rev. D36}:527-545.
\bibitem{ovlapdir} Neuberger, H. 1998.
Exactly Massless Quarks on the Lattice.
{\it Phys. Lett. B417}:141-4.
\bibitem{vectorprd} Neuberger, H. 1998.
Vector Like Gauge Theories with Almost Massless Fermions on the Lattice.
{\it Phys. Rev. D57}:5417-33.
\bibitem{moreov} Neuberger, H. 1998.
More About Exactly Massless Quarks on the Lattice. 
{\it Phys. Lett. B427}:353-5.
\bibitem{hasen} Hasenfratz, P, Laliena, V, Niedermayer, F. 1998.
The Index Theorem in QCD with a Finite Cutoff.
{\it Phys. Lett. B427}:125-131.
\bibitem{twaichiu} Chiu, T-W, 2000.
Can Nonlocal Dirac Operators be Topologically Proper?
{\it hep-lat/0008010}.
\bibitem{dubna} Neuberger, H, 2000.
Chiral Symmetry Outside Perturbation Theory. 
{\it Nato Science Series C - Mathematical and Physical Sciences 553}:113-124. 
\bibitem{prl2} Neuberger, H. 1998.
A Practical Implementation of the Overlap-Dirac Operator.
{\it Phys. Rev. Lett. 81}:4060-62.
\bibitem{higham} Higham, N. 1994.
The Matrix Sign Function and its Relation to the Polar Decomposition.
{\it Lin. Alg. and Appl. 212/213}:3-20.
\bibitem{twopass} Neuberger, H. 1999.
Minimizing Storage in Implementations of the Overlap Lattice-Dirac Operator.
{\it Int. J. Mod. Phys. C10}:1051-8.
\bibitem{chain1} Neuberger, H. 1999.
The Overlap Lattice Dirac Operator and Dynamical Fermions.
{\it Phys. Rev. D60}:065006.
\bibitem{chain2} Narayanan, R, Neuberger, H. 2000.
An Alternative to Domain Wall Fermions.
{\it Phys. Rev. D62}:074504. 
\bibitem{scri} Edwards, RG, Heller, UM, Narayanan, R. 1998.
Spectral Flow, Chiral Condensate and Topology in Lattice QCD.
{\it Nucl. Phys. B535}:403-422.
Edwards, RG, Heller, UM, Narayanan, R. 1999.
Approach to the Continuum Limit of the Quenched Hermitian Wilson-Dirac
Operator.
{\it Phys. Rev. D60}:034502.
\bibitem{localzm} Berruto, F, Narayanan, R, Neuberger, H. 2000.
Exact Local Fermionic Zero Modes. {\it Phys. Lett. B489}:243-50. 











\end{thebibliography}
\end{document}